\documentstyle[11pt]{article}
\begin{document}

\begin{center}
{\huge {\bf Arrival time in quantum mechanics \vspace*{0.6cm}\\}} {\Large 
{V. Delgado and J. G. Muga\vspace*{.2cm}}}\\{\it Departamento de F\'\i sica
Fundamental y Experimental, \\Universidad de La Laguna, 38205-La Laguna,
Tenerife, Spain\vspace*{0.7cm}\\}
\end{center}

\begin{abstract}
A self-adjoint operator with dimensions of time is explicitly constructed,
and it is shown that its complete and orthonormal set of eigenstates can be
used to define consistently a probability distribution of the time of
arrival at a spatial point.
\end{abstract}

\vspace{1.4 cm}
\begin{center}{\large \bf {I. INTRODUCTION\vspace{.2 cm}\\}}
\end{center}

The role that time plays in quantum mechanics has always been controversial.
This is in part a consequence of the rather singular status that time
exhibits in nonrelativistic physics. In particular, time enters the
Schr\"odinger equation as an external parameter and accordingly the quantum
formalism is usually concerned with probability distributions of measurable
quantities at a definite instant of time. However, one may also ask for the
instant of time at which a certain property of a quantum system takes a
given value. In this case time has the character of a dynamical variable: It
depends on the initial state of the system and on its dynamical evolution,
and appears as an intrinsic property of the physical system under study.
Since such an instant of time is, in principle, a perfectly measurable
quantity it seems natural to try to incorporate the concept of a time
observable into the quantum formalism. However this is not an easy task. The
standard quantum formalism associates measurable quantities with
self-adjoint operators acting on the Hilbert space of physical states, and
postulates that the probability distribution of the outcomes of any
well-designed measuring apparatus can be obtained in terms of the orthogonal
spectral decomposition of the corresponding self-adjoint operator, with no
explicit dependence on the particular properties of the measuring device.
Therefore, the problem reduces, in principle, to find a suitable quantum
operator. This is usually accomplished via the correspondence principle,
starting from the corresponding classical expressions and quantizing by
using certain specific quantization rules. However, in doing so one
frequently has to face with the problem that in general there exists no
unique way to obtain a quantum operator which reduces to a given known
expression in the classical limit ($\hbar \rightarrow 0$).

Given the Hamiltonian $H(q,p)$ of a conservative classical system, expressed
in terms of canonical variables $(q,p)$, one can always make a canonical
transformation to new canonical variables $(H,T)$, where $H$ is the
Hamiltonian of the system and $T$ its conjugate variable, which satisfies
the Hamilton's equation [\ref{Raz},\ref{Ko}]

\begin{equation}
\label{echam}\frac{dT}{dt}=\{H,T\}=1 , 
\end{equation}
$\{H,T\}$ denoting the Poisson bracket of $H$ and $T$. The important point
is that the above equation clearly reflects that the canonical variable $T$
is nothing but the interval of time. Thus the next step would be to take
advantage of this desirable fact and translate the above formulation to the
quantum framework. This can be easily accomplished by means of the {\em %
canonical quantization method} [\ref{QF}], which basically states that the
classical formulation remains formally valid in the quantum domain with the
substitution of Poisson brackets by commutators, $\{H,T\}\rightarrow
1/i\hbar \, [\hat H,\hat T]$, and interpreting the dynamical variables as
self-adjoint operators in the Heisenberg picture. Then, based on the
correspondence principle and the canonical quantization method one is led to
look for a self-adjoint time operator conjugate to the Hamiltonian,

\begin{equation}
\label{ec1}[\hat H,\hat T]=i\hbar . 
\end{equation}

This commutation relation, which as can be easily verified also holds true
in the Schr\"odinger picture, has the additional desirable consequence that
it implies the uncertainty relation

\begin{equation}
\label{ec1pb}\Delta H\,\Delta T\geq 1/2\;|[\hat H,\hat T]| , 
\end{equation}
with $\Delta H$ and $\Delta T$ being the usual root-mean-square deviations
of the corresponding dynamical variables. Unfortunately no such a time
operator exists. As remarked by Pauli the existence of a self-adjoint
operator satisfying the above commutation relation is incompatible with the
semibounded character of the Hamiltonian spectrum [\ref{Pauli}].

The lack of a proper time observable has a number of consequences [\ref
{Aharo}]. In particular the time--energy uncertainty relation has remained
unclear over the time. This is so basically because contrary to what happens
with the well-known position--momentum uncertainty relation, there exists no
unique way to put in a quantitative setting what is really meant by the time
spread $\Delta T$. In fact the consequences derived from incorrect
application of the time--energy uncertainty relation have led to a great
deal of confusion.

Another related problem which remains controversial at present is that
concerned with the formal definition of traversal and tunneling times [\ref
{Lan},\ref{Rev}]. This subtle question has received considerable attention
in recent years [\ref{Bu}--\ref{VDB2}] motivated, in part by the possible
applications of tunneling in semiconductor technology. However, the
simplest problem involving time as a dynamical variable is that concerned
with the time of arrival of a free particle at a given spatial point. Such a
time constitutes a well-defined concept which at a classical level can be
extracted from the formalism by simply inverting the corresponding equations
of motion. Moreover it is a perfectly measurable quantity whose probability
distribution can, in principle, be experimentally determined within any
desirable precision. However, the standard quantum theory of measurement
does not provide any formulation which allows to infer such a probability
distribution. In fact, time ago Allcock [\ref{Allco}] argued against such a
possibility. This author claims that it is not possible to construct any
operationally meaningful and apparatus-independent probability formula. Even
though more recently a number of works addressing this question from a more
optimistic perspective have appeared [\ref{Du},\ref{Kij}--\ref{Jleon}], the
problem is yet far from being resolved, and additional investigation on this
fundamental question is worthwhile.

In this paper we analyze the possibility of defining a probability
distribution for the arrival time of a quantum particle at a definite
spatial point. Specifically, we are interested in searching for an
apparatus-independent theoretical prediction for the probability
distribution of arrival times at a given spatial point, as a certain
function of the initial state of the system.
Our results turn out to be similar to those previously obtained by 
Kijowski [\ref{Kij}]. However, the approach by Kijowski is based on 
the definition of a non-conventional wave function which evolves on a 
family of $x \!\! = \!\! {\rm const}$ planes (instead of evolving in 
time according to the Schr\"odinger equation), and whose relation to the 
conventional wave function is unclear [\ref{Mielnik},\ref{Piron}]. 
Our approach, on the contrary,
is entirely developed within the formalism of standard quantum mechanics.

We begin considering in detail the case of a free quantum particle and then
we study the more interesting case of a quantum particle scattered by a
potential barrier. For simplicity we shall restrict ourselves to a one
spatial dimension. For our purpose, it proves to be useful analyzing first
in some detail the reason for the nonexistence of a self-adjoint time
operator in quantum mechanics.

\vspace{1.2 cm}
\begin{center}
{\large \bf {II. NONEXISTENCE OF A TIME OPERATOR IN \\ QUANTUM MECHANICS
\vspace{.2 cm}\\}}
\end{center}

As stated above, according to Pauli's argument, because of the semibounded
character of the energy spectrum there exists no self-adjoint operator
conjugate to the Hamiltonian, i.e., satisfying the commutation relation (\ref
{ec1}). The same negative conclusion was found by Allcock [\ref{Allco}]
using a somewhat different argument based on the time-translation property
of the arrival time concept.

If $\{|T\rangle \}$ denotes a set of measurement eigenstates for the arrival
time at a given spatial point of a particle in the quantum state $|\psi
\rangle $, then according to the standard quantum formalism the probability
amplitude for the arrival time at the instant $t=T$, would be given by $\psi
(T)=\langle T|\psi \rangle $. If one translates the state of the system
forward through time by an amount $\tau $, i.e., $|\psi \rangle \rightarrow
|\psi ^{\prime }\rangle =\exp (-i\hat H\tau /\hbar )|\psi \rangle $, then it
seems natural to expect the probability amplitude to transform according to $%
\psi (T)\rightarrow \psi ^{\prime }(T)=\psi (T+\tau )$. Since this
transformation property must be true for any state vector $|\psi \rangle $
it follows that the measurement eigenstates $\{|T\rangle \}$ must satisfy 
\begin{equation}
\label{ec2}|T+\tau \rangle =e^{i\hat H\tau /\hbar } \, |T\rangle , 
\end{equation}
which reflects the fact that under a translation backward in time by an
amount $\tau $, any measurement eigenstate corresponding to arrival time at
the instant $t=T$, transforms into another measurement eigenstate,
corresponding to an arrival time $t=T+\tau $. Based on general grounds,
Allcock showed that measurements eigenstates with such a desirable property
cannot be orthogonal, which implies that it is not possible to construct the
corresponding self-adjoint arrival-time operator. It is not difficult to see
that this negative conclusion can be traced back again to the semi-infinite
nature of the Hamiltonian spectrum. To this end let us consider the
following three statements:

\begin{enumerate}
\item[i)]  There exists a self-adjoint operator $\hat T$ conjugate to the
Hamiltonian $\hat H$, i.e., satisfying $[\hat H,\hat T]=i\hbar $.

\item[ii)]  There exists a self-adjoint operator $\hat T$, whose
(orthonormal and complete) set of eigenstates $\{|T\rangle \}$ transforms
under time-translations as $e^{i\hat H\tau /\hbar }\,|T\rangle =|T+\tau
\rangle $.

\item[iii)]  There exists a self-adjoint operator $\hat T$ which generates
unitary energy-translations, i.e., such that for any energy eigenstate $%
|E\rangle $ and any parameter $\varepsilon $ with dimensions of energy, it
holds 
\begin{equation}
\label{ec3}e^{i\hat T\varepsilon /\hbar }\,|E\rangle =|E-\varepsilon 
\rangle , 
\end{equation}
where the operator $\hat T$ is assumed to be defined onto the whole Hilbert
space spanned by the Hamiltonian eigenstates.
\end{enumerate}

It is not difficult to see that these statements are in fact equivalent.
Indeed, if i) is true, then by induction, one has

\begin{equation}
\label{ind}[\hat H^n,\hat T]=in\hbar \hat H^{n-1},\;\;\;\;\;\;n\geq 1 ,
\end{equation}
where $\hat H^0\equiv {\bf 1}$. Of course the validity of Eq. (\ref{ind})
rests on the reasonable assumption that the Hamiltonian is well-behaved
enough as to guarantee the existence of all its higher integer powers. Since
it also holds that $[\hat H^n,\hat T]=0$ for $n=0$, then multiplying Eq. (%
\ref{ind}) by $\left( i\tau /\hbar \right) ^n/n!$ ($\tau $ being an
arbitrary parameter with dimensions of time) and summing from $n=0$ to $%
n=\infty $, one finds

\begin{equation}
\label{ec4}[e^{i\hat H\tau /\hbar },\hat T]=-\tau \,e^{i\hat H\tau /\hbar }
. 
\end{equation}
If $\{|T\rangle \}$ denotes a complete and orthonormal set of eigenstates of 
$\hat T$, then according to Eq. (\ref{ec4}) it holds 
\begin{equation}
\label{ec5}\hat T \, e^{i\hat H\tau /\hbar }\,|T\rangle =(T+\tau )\,
e^{i\hat H\tau /\hbar }\,|T\rangle , 
\end{equation}
which after suitable choice of normalization and phase leads to statement
ii). Conversely, if ii) is true for any eigenstate $|T\rangle $ and any
parameter $\tau $, then one can repeat the same steps backward to reach i).

On the other hand, it can be readily seen that statement i) also implies
statement iii). Indeed, if i) holds one has by induction that

\begin{equation}
\label{ec6}[\hat H,\hat T^n]=in\hbar \hat T^{n-1},\;\;\;\;\;\;n\geq 0 ,
\end{equation}
($\hat T^0\equiv {\bf 1}$), which implies that for any parameter $%
\varepsilon $ with dimensions of energy

\begin{equation}
\label{ec7}[\hat H,e^{i\hat T\varepsilon /\hbar }]=-\varepsilon \,e^{i\hat
T\varepsilon /\hbar }. 
\end{equation}
Therefore, according to Eq. (\ref{ec7}) any energy eigenstate $|E\rangle $
verifies

\begin{equation}
\label{ec8}\hat{H\,}e^{i\hat{T}\varepsilon /\hbar }\,|E\rangle
=(E-\varepsilon )\,e^{i\hat{T}\varepsilon /\hbar }\,|E\rangle , 
\end{equation}
from which after proper normalization follows iii). An analogous reasoning
can be repeated from iii) to i), which shows the equivalence among the above
three statements.

Since iii) is obviously incompatible with a semibounded Hamiltonian
spectrum, it follows that it is not possible to find a self-adjoint arrival
time operator satisfying the desirable conditions i) or ii).

\vspace{1.2 cm}
\begin{center}
{\large \bf 
{III. A SELF-ADJOINT OPERATOR WITH \\ DIMENSIONS OF TIME\vspace{.2 cm}\\}}
\end{center}

We start by considering the simplest conceivable arrival time problem,
namely a one-dimensional free particle moving along the $x$-axis towards a
detector. In looking for a probability distribution of the time of arrival
it is most convenient to work in the energy representation $\{|E,\alpha
\rangle ;\;E\geq 0,\;\alpha =+,-\}$, defined by the eigenvalue equations 
\begin{equation}
\label{ec9}\hat H_0\,|E,\pm \rangle =E\,|E,\pm \rangle , 
\end{equation}
\begin{equation}
\label{ec10}\hat P\,|E,\pm \rangle =\pm \sqrt{2mE}\,|E,\pm \rangle , 
\end{equation}
where $\hat H_0\!=\!\hat P^2/2m$ is the Hamiltonian of the free particle,
and $\hat P$ its momentum operator. The orthonormal and complete set of
energy eigenstates $\left\{ |E,\alpha \rangle \right\} $, which satisfy

\begin{equation}
\label{res1}\sum_{\alpha =\pm }\int_0^\infty dE\,|E,\alpha \rangle \langle
E,\alpha |={\bf 1} , 
\end{equation}

\begin{equation}
\label{ort1}\langle E,\alpha |E^{\prime },\alpha ^{\prime }\rangle =\delta
_{\alpha \alpha ^{\prime }}\,\delta (E-E^{\prime }) , 
\end{equation}
can be expressed in terms of the usual momentum representation by means of
the relation

\begin{equation}
\label{ec11}|E,\pm \rangle =(m/2E)^{1/4}\,|p=\pm \sqrt{2mE}\rangle , 
\end{equation}
where the momentum eigenstates $\left\{ \mid p\rangle \right\} $ are
normalized as

\begin{equation}
\label{ec12}\int_{-\infty}^{+\infty} dp\,|p\rangle \langle p|={\bf 1}, 
\end{equation}

\begin{equation}
\label{ec13}\langle p|p^{\prime }\rangle =\delta (p-p^{\prime }) . 
\end{equation}

As stated above, the impossibility of finding a time-of-arrival operator can
always be traced back to the bounded character of the Hamiltonian spectrum.
To circumvent such a difficulty we shall instead look for a self-adjoint
operator $\hat {{\cal T}}$ with dimensions of time, conjugate to a
conveniently defined self-adjoint operator $\hat {{\cal H}}$, with
dimensions of energy and a non-bounded spectrum,

\begin{equation}
\label{ec14}[\hat{{\cal H}},\hat{{\cal T}}]=i\hbar . 
\end{equation}

Of course this is a somewhat arbitrary procedure, since the definition of $%
\hat {{\cal T}}$ depends in a fundamental way on the arbitrary choice one
makes for the operator $\hat {{\cal H}}$. Moreover, as long as $\hat {{\cal H%
}}$ differs from the Hamiltonian of the system, the corresponding operator $%
\hat {{\cal T}}$ could not be associated to the actual physical time.
Therefore, it remains the fundamental question of verifying whether it is
possible to give a proper physical interpretation to the selected $\hat {%
{\cal T}}$-operator in terms of measurements results, i.e., whether it is
possible to define an algorithm which enables us to connect the probability
distribution of measurement results with the set of eigenvalues and
eigenstates of the operator $\hat {{\cal T}}$. At this point we shall
postpone this essential question and simply consider the procedure just
outlined to be worth exploring.

We now introduce projectors, $\Theta (\pm \hat{P})$, onto the subspaces
generated by plane waves with positive/negative momenta,

\begin{equation}
\label{ec16}\Theta (\pm \hat P)=\!\int_0^\infty dp\mid \!\pm p\rangle
\langle \pm p \! \mid , 
\end{equation}
and define the self-adjoint operator 
\begin{equation}
\label{ec15}{\rm sgn}(\hat P)\equiv \Theta (\hat P)-\Theta (-\hat P) . 
\end{equation}
Obviously, ${\rm sgn}(\hat P)$ commutes with the Hamiltonian and satisfies
the eigenvalue equation 
\begin{equation}
\label{ec17}{\rm sgn}(\hat P)\,|E,\pm \rangle =\pm \,|E,\pm \rangle . 
\end{equation}
This operator allows us to define a simple self-adjoint operator with
dimensions of energy, 
\begin{equation}
\label{ec18}\hat {{\cal H}}\equiv {\rm sgn}(\hat P)\,\hat H_0 , 
\end{equation}
which exhibits a non-bounded spectrum,

\begin{equation}
\label{ec19}\hat{{\cal H}}\,|E,\pm \rangle =\pm \,E\,|E,\pm \rangle ,
\;\;\;\;\;\;\;(E\geq 0) . 
\end{equation}

Notice that this is, in a sense, the simplest choice, since the restrictions
of $\hat {{\cal H}}$ to the subspaces spanned by plane waves with 
positive/negative momentum coincide with plus/minus the corresponding 
restrictions of the Hamiltonian $\hat H_0$. Specifically, 
\begin{equation}
\label{ec20}\Theta (\pm \hat P)\,\hat {{\cal H}}\,\Theta (\pm \hat P)=\pm
\,\Theta (\pm \hat P)\,\hat H_0 \, \Theta (\pm \hat P) . 
\end{equation}

Introducing the following notation for the energy eigenstates

\begin{equation}
\label{ec21}|\varepsilon \rangle = \left\{ 
\begin{array}{ll}
|\!+ \! E\rangle \equiv |E,+ \rangle & \mbox{ if  $\varepsilon \geq 0$} \\ 
|\!- \! E\rangle \equiv |E,- \rangle & \mbox{ if  $\varepsilon < 0$} 
\end{array}
\right. 
\end{equation}
the above results can be rewritten in terms of the complete and orthonormal
set of states $\{|\varepsilon \rangle ;\;\varepsilon \in (-\infty ,+\infty
)\}$ satisfying the eigenvalue equations

\begin{equation}
\label{ec22}\hat {{\cal H}}\,|\varepsilon \rangle =\varepsilon
\,|\varepsilon \rangle , 
\end{equation}

\begin{equation}
\label{ec23}\hat{P}\,|\varepsilon \rangle ={\rm sgn}(\varepsilon )\sqrt{%
2m|\varepsilon |}\,|\varepsilon \rangle , 
\end{equation}

\begin{equation}
\label{ec24}\hat H_0 \,|\varepsilon \rangle =|\varepsilon |\,|\varepsilon
\rangle . 
\end{equation}

Now searching for a self-adjoint operator $\hat{{\cal T}}$ conjugate to $%
\hat{{\cal H}}$ is a straightforward matter. To this end let us introduce
the states $|\tau \rangle $ defined in the $\{\,|\varepsilon \rangle \}$
basis as

\begin{equation}
\label{ec25}|\tau \rangle =h^{-1/2}\!\int_{-\infty }^{+\infty }\!
d\varepsilon\,e^{i\varepsilon \tau /\hbar }\,|\varepsilon \rangle . 
\end{equation}
These states also constitute a complete and orthonormal set. Indeed,

\begin{equation}
\label{ec27}\langle \tau |\tau ^{\prime }\rangle =\!\! \int_{-\infty
}^{+\infty}\! d\varepsilon \,\langle \tau |\varepsilon \rangle \langle
\varepsilon |\tau ^{\prime }\rangle =h^{-1}\!\int_{-\infty }^{+\infty
}\!d\varepsilon \,e^{-i\varepsilon (\tau -\tau ^{\prime })/\hbar }=\delta
(\tau -\tau ^{\prime }) , 
\end{equation}

\begin{eqnarray}
\label{ec28}\!\int_{-\infty }^{+\infty }\!d\tau \,|\tau 
\rangle \langle \tau
|\!\!&=&\!\!h^{-1}\!\int_{-\infty }^{+\infty }\!d\tau\! \int_{-\infty }^{+
\infty}\!
d\varepsilon \!\int_{-\infty }^{+\infty }\!d\varepsilon ^{\prime
}\,e^{i(\varepsilon -\varepsilon ^{\prime })\tau /\hbar }\,|\varepsilon
\rangle \langle \varepsilon ^{\prime }| \nonumber \\ 
&=&\!\!\!\int_{-\infty }^{+\infty
}\!d\varepsilon \,|\varepsilon \rangle \langle \varepsilon |={\bf 1} .
\end{eqnarray}
We can therefore define a self-adjoint operator, with eigenstates and
eigenvalues given by $|\tau \rangle $ and $\tau $, respectively, in terms of
its spectral decomposition,

\begin{equation}
\label{ec29}\hat{{\cal T}}=\!\int_{-\infty }^{+\infty }\!d\tau \,\tau
\,|\tau\rangle \langle \tau | . 
\end{equation}
The operator so defined has dimensions of time and automatically satisfies
the commutation relation (\ref{ec14}). However, there exists no guarantee
that it will be useful in the time-of-arrival problem. In fact, $\hat {{\cal %
T}}$ turns out to be invariant under time reversal, and consequently the
variable $\tau $ cannot be identified with the physical time $t$. This can
be most easily seen in the momentum representation,

\begin{equation}
\label{ec29b}\langle p|\tau \rangle =(|p|/mh)^{1/2}\,e^{i\,{\rm sgn}(p)\frac{%
p^2}{2m}\tau /\hbar } . 
\end{equation}
Let $\hat {{\cal R}}$ denote the time-reversal operator, then we have

\begin{equation}
\label{ec29c}\hat {{\cal R}}\,|\tau \rangle =\!\int_{-\infty }^{+\infty
}\!dp \mid\!\! -p\rangle \langle p|\tau \rangle ^{*}=\!\int_{-\infty }^{+
\infty}\!dp\mid\!\! -p \rangle \langle -p|\tau \rangle =|\tau \rangle , 
\end{equation}
so that, according to Eq. (\ref{ec29}), it holds that $\hat {{\cal R}}\,\hat
{{\cal T}}\,\hat {{\cal R}}^{\dagger }=\hat {{\cal T}}$. Moreover,

\begin{equation}
\label{ec29d}|\tau +\tau ^{\prime }\rangle =e^{i\hat {{\cal H}}\tau ^{\prime 
}/\hbar }\,|\tau \rangle \neq e^{i\hat H_0\tau ^{\prime }/\hbar }\,|\tau
\rangle , 
\end{equation}
and the states $\{\,|\tau \rangle \}$ do not exhibit the desirable
time-translation property (\ref{ec2}) either.

In spite of these facts, it is possible to give a physical interpretation to
the operator $\hat {{\cal T}}$. As we shall see below, one can consistently
define a probability distribution of arrival times in terms of the
eigenvalues and eigenstates of this operator. To this end it is convenient
to decompose $|\tau \rangle $ as a superposition of negative- and
positive-momentum contributions,

\begin{equation}
\label{ec34}|\tau \rangle =h^{-1/2}\!\int_0^{\infty }\!dE\, e^{-iE\tau
/\hbar}\,|E,-\rangle +h^{-1/2}\!\int_0^{\infty }\!dE\, e^{iE\tau /\hbar
}\,|E,+\rangle . 
\end{equation}

Defining new states

\begin{equation}
\label{ec30}|t,\pm \rangle \equiv h^{-1/2}\!\int_0^{\infty }\!dE\,
e^{iEt/\hbar }\,|E,\pm\rangle , 
\end{equation}
we see that $|\tau \rangle $ can be written in the form

\begin{equation}
\label{ec34b}\,|\tau \rangle=|t\!=\!-\tau,-\rangle+|t\!=\!+\tau ,+\rangle .
\end{equation}

The important point is that $|\tau \rangle $ has been decomposed in terms of
states $\left\{ |t,\pm \rangle \right\} $ which do satisfy the
time-translation property (\ref{ec2}) 
\begin{equation}
\label{ec30b}|t+\tau ^{\prime },\pm \rangle =e^{i\hat H_0\tau ^{\prime
}/\hbar }\,|t,\pm \rangle , 
\end{equation}
and transform under time reversal as

\begin{equation}
\label{ec30c}\hat {{\cal R}}\,|t,\pm \rangle =|-t,\mp \rangle ,
\end{equation}
so that the variable $t$, unlike $\tau $, could, in principle, be associated
with physical time.

Note that even though the states $\left\{ |t,\pm \rangle \right\} $
constitute a complete set,

\begin{equation}
\label{ec31}\sum_{\alpha =\pm }\int_{-\infty }^{+\infty}dt\,|t,\alpha\rangle
\langle t,\alpha |={\bf 1} , 
\end{equation}
they are not orthogonal,

\begin{eqnarray}
\label{ec32}
\langle t,\alpha |t^{\prime },\alpha ^{\prime }\rangle\!\! &=&\!\!\!
\sum_{\beta =\pm
}\int_0^{\infty }\!dE\,\langle t,\alpha |E,\beta \rangle \langle E,\beta
|t^{\prime },\alpha ^{\prime }\rangle =\frac{\delta _{\alpha \alpha ^{\prime
}}}h \!\int_0^{\infty }\!dE\,e^{-iE(t-t^{\prime })/\hbar } \nonumber \\
&=&\!\!\!\frac 12\delta _{\alpha \alpha ^{\prime }}\left\{ \delta
(t-t^{\prime })-P.P.\frac i{\pi (t-t^{\prime })}\right\} .
\end{eqnarray}

For this reason, the states $\left\{ |t,\pm \rangle \right\} $ cannot be
used to construct a self-adjoint operator.

\vspace{1.2 cm}
\begin{center}{\large \bf 
{IV. MEAN ARRIVAL TIME\vspace{.2 cm}\\}}
\end{center}

One of the most controversial aspects of quantum mechanics is that
concerning with the connection between the theoretical formulation and the
corresponding measurement results. In its space-time representation, quantum
mechanics becomes a continuous wave theory, whereas measuring devices
usually deal with individual particles. The quantum formalism tells us how
to obtain the probability distribution of the measurement results in terms
of projections of the state vector onto appropriate subspaces of the Hilbert
space. While in the standard interpretation it is commonly assumed that
probability distributions refer to individual particles, their experimental
verification requires an ensemble. Quantities defined in the ensemble may
offer practical guidance not only in the interpretation of quantum
measurement theory, but also in the search for the quantum counterpart of a
classical physical variable. In this sense, the mean arrival time may be
useful in looking for a probability distribution of arrival times.

Consider a classical statistical ensemble of particles of mass $m$, directed
along a well-defined direction, and characterized by the phase space
distribution function $f(x,p,t)$. The average time of arrival at a spatial
point $x_0$ is given by

\begin{equation}
\label{tav}\langle t_{x_0}\rangle =\frac{\int_{-\infty }^{+\infty
}dt\,t\,J(x_0,t)}{\int_{-\infty }^{+\infty }dt\,J(x_0,t)} , 
\end{equation}
where $J(x_0,t)$ represents the average current at $x_0$,

\begin{equation}
\label{jota}J(x_0,t)=\!\int\!\! \int\! f(x,p,t)\,\frac pm \, \delta
(x-x_0)\,dx\,dp , 
\end{equation}
and plays the role of an unnormalized probability distribution of arrival
times. It seems natural to make use of the correspondence principle in order
to translate the expression for the classical average time of arrival, Eq. (%
\ref{tav}), to the quantum formalism. This can be accomplished by
substituting $J(x_0,t)$ by the expectation value of the current operator

\begin{equation}
\label{curop}\hat{J}(X)=\frac 1{2m}\left( \hat{P}\,|X\rangle \langle
X|+|X\rangle \langle X|\,\hat{P}\right) .
\end{equation}

Such a quantum definition for the average time of arrival has been widely
used in recent years [\ref{Du},\ref{Leav2}--\ref{Mug}]. However, unlike its
classical counterpart, even for wave packets directed along a well-defined
spatial direction, the quantum probability current is not positive definite.
For this reason, it cannot be considered as a probability distribution of
individual arrival times, and the validity of the above expressions in the
quantum context is questionable. In fact, strictly speaking, $\hat J(X)$ is
an operator valued distribution (the operator analog of a generalized
function), and as pointed out by Wigner [\ref{Wig2}], there exist quantum
quantities, such as $\hat J(X)$, whose expectation values do not correspond
to averages of individual measurements (eigenvalues), but represent a
measurable property of the ensemble as a whole.

In spite of the general inadequacy of $\hat J(X)$ to describe the
probability distribution of arrival times, when quantum backflow
contributions become negligible the quantum current becomes positive and
admits a probability interpretation [\ref{Leav2}]. Such a situation occurs,
at large times, for freely moving packets containing only positive momenta,
and it also occurs under the standard asymptotic conditions of scattering
theory. In fact, Eq. (\ref{tav}) can be operationally justified in the
quantum case by using a {\em perfect absorber}, i.e., a complex potential
that absorbs the incoming wave in an arbitrary small spatial region, without
reflection or transmission [\ref{Mug}]. According to such an operational
model, which simulates the detection of incoming particles by a destructive
procedure, the average time given by (\ref{tav}) coincides with the average
time of absorption (detection) within any desirable precision. Thus, any
properly defined arrival time probability distribution should be compatible
with Eq. (\ref{tav}).

\vspace{1.2 cm}
\begin{center}
{\large \bf {V. PROBABILITY DISTRIBUTION OF ARRIVAL TIMES.\\ FREE PARTICLE
\vspace{.2 cm}\\}}
\end{center}

We shall restrict ourselves to the case of a free particle moving along a
well-defined direction towards a detector situated at the point $x=X$.
Specifically, we assume that the ingoing asymptote of the actual state of
the particle corresponds, in the position representation, to a wave packet
which is either a linear superposition of positive plane waves or a linear
superposition of negative plane waves,

\begin{equation}
\label{psin}|\psi _{\pm ,{\rm in}}\rangle \equiv \Theta (\pm \hat P)\,|\psi
_{\pm ,{\rm in}}\rangle . 
\end{equation}
Under these circumstances, the in asymptote becomes indistinguishable from
the actual state $|\psi _{\pm }(t\!=\!0)\rangle $, so that we shall not
discriminate between them from now on. Note that as a consequence of the
commutation between the time-evolution operator $e^{-i\hat H_0 t/\hbar }$
and the projectors $\Theta (\pm \hat P)$, at any time $t$ it also holds

\begin{equation}
\label{psit}|\psi _{\pm }(t)\rangle \equiv \Theta (\pm \hat{P})\,|\psi _{\pm
}(t)\rangle . 
\end{equation}

As stated in the previous section, the mean arrival time at a point $X$ is
given by

\begin{equation}
\label{qtav}\langle t_X\rangle _{\pm }=\frac{\int_{-\infty }^{+\infty }d\tau
\,\tau \,\langle \psi _{\pm }(\tau )|\hat J(X)|\psi _{\pm }(\tau )\rangle }{%
\int_{-\infty }^{+\infty }d\tau \,\langle \psi _{\pm }(\tau )|\hat J(X)|\psi
_{\pm }(\tau )\rangle } , 
\end{equation}
where $\hat J(X)$ is the current operator in the Schr\"odinger picture,
given in Eq. (\ref{curop}), and $\langle \psi _{\pm }(\tau )|\hat J(X)|\psi
_{\pm }(\tau )\rangle $, is the probability current at the instant of time $%
t\equiv \tau $, in the (Schr\"odinger) state $|\psi _{\pm }(\tau )\rangle $.

In the free case, we have

\begin{equation}
\label{ec37b}\int_{-\infty }^{+\infty }\!d\tau \,\langle \psi _{\pm }(\tau
)|\hat J(X)|\psi _{\pm }(\tau )\rangle =\pm 1 , 
\end{equation}
so that, the mean arrival time $\langle t_X\rangle _{\pm }$ can be expressed
as the following expectation value

\begin{equation}
\label{ec41}\langle t_X\rangle _{\pm }\equiv \pm \langle \psi _{\pm }|\hat {%
{\cal J}}_{\pm }(X)|\psi _{\pm }\rangle , 
\end{equation}
where $|\psi _{\pm }\rangle $ denotes the state of the particle in the
Heisenberg picture, i.e.,

\begin{equation}
\label{ec39}|\psi _{\pm }\rangle =e^{i\hat H_0 \tau /\hbar }\,|\psi _{\pm
}(\tau )\rangle =|\psi _{\pm }(0)\rangle , 
\end{equation}
and we have introduced the operator

\begin{equation}
\label{ec42}\hat {{\cal J}}_{\pm }(X)\equiv \!\int_{-\infty }^{+\infty }
\!d\tau \,\tau \,\Theta (\pm \hat P)\, \hat J_{{\rm H}}(X,\tau )\,\Theta
(\pm \hat P) , 
\end{equation}
where $\hat J_{{\rm H}}(X,\tau )$ is the Heisenberg current operator,

\begin{equation}
\label{ec40}\hat J_{{\rm H}}(X,\tau )=e^{i\hat H_0\tau /\hbar }\,\hat
J(X)\,e^{-i\hat H_0\tau /\hbar }. 
\end{equation}
For later convenience use has been made in the above equations of the
identity $|\psi _{\pm }(0)\rangle \equiv \Theta (\pm \hat P)\,|\psi _{\pm
}(0)\rangle $.

Inserting twice the resolution of unity, Eq. (\ref{res1}), and using 
\begin{equation}
\label{ec43}\Theta (\pm \hat{P})\,|E,\alpha \rangle =\delta _{\alpha ,\pm
}\,|E,\alpha \rangle , 
\end{equation}
$\hat {{\cal J}}_{\pm }(X)$ takes the form

\begin{equation}
\label{ec44}\hat {{\cal J}}_{\pm }(X)=\!\!\int_{-\infty }^{+\infty }\!\!
d\tau \,\tau\!\!\int_0^{\infty }\!\!dE\!\!\int_0^{\infty }\!\!dE^{\prime}
e^{i(E-E^{\prime })\tau /\hbar }\,|E,\pm \rangle \langle E,\pm |\hat
J(X)|E^{\prime },\pm \rangle \langle E^{\prime },\pm | . 
\end{equation}
Substituting the expression (\ref{curop}) for $\hat J(X)$, using Eq. (\ref
{ec10}), and taking into account that according to Eq. (\ref{ec11})

\begin{equation}
\label{ec45}\langle X|E,\pm \rangle =h^{-1/2}(m/2E)^{1/4}\,e^{\pm i\sqrt{2mE}%
\,X/\hbar } , 
\end{equation}
the matrix element in the integrand of (\ref{ec44}) can be rewritten in the
form

\begin{equation}
\label{ec46}\langle E,\pm |\hat J(X)|E^{\prime },\pm \rangle =\pm \frac
1{2h}\!\left\{\! \left( \frac E{E^{\prime }}\right)^{\!1/4}\!+\!\left( \frac{%
E^{\prime }}E\right)^{\!1/4}\right\} e^{\mp i\left( \sqrt{2mE}- \sqrt{%
2mE^{\prime }}\right) X/\hbar } . 
\end{equation}
After insertion of Eq. (\ref{ec46}) into Eq. (\ref{ec44}), the operator $%
\hat {{\cal J}}_{\pm }(X)$ reads

\begin{equation}
\label{ec47}\hat {{\cal J}}_{\pm }(X)=e^{-i\hat PX/\hbar }\,\hat {{\cal J}%
}_{\pm }(X\!=\!0)\,e^{+i\hat PX/\hbar } , 
\end{equation}
where we have again taken advantage of Eq. (\ref{ec10}) to write

\begin{equation}
\label{ec47b}e^{i\hat PX/\hbar }\,|E,\pm \rangle = e^{\pm i\sqrt{2mE}\,
X/\hbar }\,|E,\pm \rangle , 
\end{equation}
and $\hat {{\cal J}}_{\pm }(X\!=\!0)$, which is the operator involved in the
determination of the mean arrival time at the point $X=0$, is given by

\begin{equation}
\label{ec48}\hat {{\cal J}}_{\pm }(0)=\pm \!\int_{-\infty }^{+\infty }
\!\!d\tau\,\tau\!\! \int_0^{\infty }\!\!dE\!\!\int_0^{\infty }\!\!
dE^{\prime } \frac 1{2h}\!\left\{\! \left( \frac E{E^{\prime }}
\right)^{\!1/4}\!+\!\left(\frac{E^{\prime }}E\right)^{\!1/4}\right\}
e^{i(E-E^{\prime })\tau /\hbar }\,|E,\pm \rangle \langle E^{\prime },\pm | . 
\end{equation}

In order to guarantee that the integrand is well behaved over the whole
interval of integration we shall restrict ourselves to physical states
satisfying the boundary conditions

\begin{equation}
\label{ec49}{\rm \lim _{E\rightarrow \infty }\,}E^{1/4}\,\langle E,\pm |\psi
_{\pm }\rangle =0 , 
\;\;\;\;\;\;\;\;\;\; 
{\rm \lim _{E\rightarrow 0}\,}E^{-1/4}\,\langle E,\pm |\psi _{\pm }\rangle
=0 , 
\end{equation}
which, in the more familiar momentum representation, take the form

\begin{equation}
\label{ec51}{\rm \lim _{p\rightarrow \pm \infty }\;}\langle p|\psi _{\pm
}\rangle =0,
\;\;\;\;\;\;\;\;\;\;
{\rm \lim _{p\rightarrow 0}\;\,}p^{-1}\,\langle p|\psi _{\pm }\rangle =0, 
\end{equation}
Put another way, we shall restrict ourselves to normalizable wave packets,
which are superposition of either positive or negative plane waves, and
vanishing faster than $p$ as $p$ approaches zero.

The integral in the $\tau $ variable, in Eq. (\ref{ec48}), can be readily
performed to obtain

\begin{equation}
\label{ec53}\hat {{\cal J}}_{\pm }(0)=\mp \frac{i\hbar }2 \! \int_0^{\infty
}\!\!dE\!\!\int_0^{\infty }\!\! dE^{\prime } \!\left( \frac \partial
{\partial E}\delta (E-E^{\prime })\right) \! \left\{\! \left( \frac
E{E^{\prime }} \right)^{\!1/4}\!+\!\left(\frac{E^{\prime }}%
E\right)^{\!1/4}\right\} |E,\pm \rangle \langle E^{\prime },\pm | . 
\end{equation}

To proceed further it is convenient to consider the matrix elements of $\hat
{{\cal J}}_{\pm }(0)$ between arbitrary states $|\Phi \rangle $, $|\Psi
\rangle $ satisfying the boundary conditions (\ref{ec49}). Using the
derivative of the Dirac delta in the integrand of (\ref{ec53}) to perform
one of the two energy integrals, one arrives at

\begin{equation}
\label{ec54}\langle \Phi |\hat{{\cal J}}_{\pm }(0)|\Psi \rangle =\pm i\hbar
\!\int_0^{\infty }\!dE\,\langle E,\pm |\Psi \rangle \,\frac \partial
{\partial E}\langle \Phi |E,\pm \rangle . 
\end{equation}
On the other hand, using the resolution of the unity, Eq. (\ref{ec28}), and
taking into account that

\begin{equation}
\label{ec55}\pm i\hbar \frac \partial {\partial E}\langle \tau |E,\pm
\rangle =\tau \,\langle \tau |E,\pm \rangle , 
\end{equation}
one can obtain a useful alternative expression for the energy derivative in
the integrand of Eq. (\ref{ec54}), in terms of the self-adjoint ''time''
operator defined in Eq. (\ref{ec29}). Indeed,

\begin{eqnarray}
\label{ec56}\pm i\hbar \frac \partial {\partial E}\langle \Phi |E,\pm
\rangle \!&=&\!\pm i\hbar \!\int_{-\infty }^{+\infty }\!d\tau \,
\langle \Phi |\tau\rangle \,\frac \partial {\partial E}\langle \tau 
|E,\pm \rangle \nonumber\\
&=& \!\!\int_{-\infty }^{+\infty }\!d\tau \,\tau \,\langle \Phi |\tau
\rangle \langle \tau |E,\pm \rangle =\langle \Phi |\hat{{\cal T}}|E,\pm
\rangle .
\end{eqnarray}
Therefore,

\begin{equation}
\label{ec58}\langle \Phi |\hat{{\cal J}}_{\pm }(0)|\Psi \rangle
=\!\int_0^{\infty }\!dE\,\langle \Phi |\hat{{\cal T}}|E,\pm \rangle \langle
E,\pm |\Psi \rangle . 
\end{equation}
Taking $|\Phi \rangle =|\Psi \rangle =$ $e^{+i\hat{P}X/\hbar }|\psi _{\pm
}\rangle $, we have

\begin{eqnarray}
\label{ec59}\langle t_X\rangle _{\pm }&\equiv& \pm \langle \psi _{\pm
}|e^{-i\hat PX/\hbar }\,\hat {{\cal J}}_{\pm }(0)\,e^{+i\hat PX/\hbar
}|\psi _{\pm }\rangle \nonumber \\
&=& \pm \!\int_0^{\infty }\!dE\,\langle \psi _{\pm }|e^{-i\hat
PX/\hbar }\,\hat {{\cal T\,}}|E,\pm \rangle \langle E,\pm |e^{+i\hat
PX/\hbar }|\psi _{\pm }\rangle .
\end{eqnarray}
Using the identity

\begin{equation}
\label{ec61}|E,\pm \rangle \langle E,\pm|\equiv\Theta (\pm \hat P)
\sum_{\alpha=\pm }|E,\alpha \rangle \langle E,\alpha | , 
\end{equation}
as well as Eq. (\ref{res1}), and taking into account that $\Theta (\pm \hat
P)\,e^{+i\hat PX/\hbar }|\psi _{\pm }\rangle =e^{+i\hat PX/\hbar }|\psi
_{\pm }\rangle $, we finally find

\begin{equation}
\label{ec62}\langle t_X\rangle _{\pm }=\pm \langle \psi _{\pm }|e^{-i\hat
PX/\hbar }\,\hat {{\cal T}}\,e^{+i\hat PX/\hbar }|\psi _{\pm }\rangle . 
\end{equation}

Accordingly, the self-adjoint operator involved in the determination of the
mean arrival time at an arbitrary point $X$, is given by the spatial
translation of the operator $\hat{{\cal T}}\,$ previously defined,

\begin{equation}
\label{ec67}\hat {{\cal T}}(X)=e^{-i\hat PX/\hbar }\,\hat {{\cal T}%
}\,e^{+i\hat PX/\hbar } , 
\end{equation}
and its spectral resolution reads

\begin{equation}
\label{ec68}\hat {{\cal T}}(X)=\!\int_{-\infty }^{+\infty }\!d\tau \,\tau
\,|\tau ;X\rangle \langle \tau ;X| , 
\end{equation}
where

\begin{equation}
\label{ec69}|\tau ;X\rangle =e^{-i\hat PX/\hbar }\,|\tau \rangle
=h^{-1/2}\!\int_{-\infty }^{+\infty }\!d\varepsilon \, e^{i(\varepsilon \tau
-{\rm sgn}(\varepsilon )\sqrt{2m\left| \varepsilon \right| }\,X)/\hbar
}\,|\varepsilon \rangle . 
\end{equation}

Since the states $\{|\tau ;X\rangle \}$ are generated from the complete and
orthonormal set $\{|\tau \rangle \}$ via a unitary transformation, they also
constitute, for a given $X$, a complete and orthonormal set.

Introducing the complete but nonorthogonal set of shifted states $\{|t,\pm
;X\rangle \}$, defined as the spatial translation of the set $\{|t,\pm
\rangle \}$,

\begin{equation}
\label{ec70}|t,\pm ;X\rangle \equiv e^{-i\hat PX/\hbar }\,|t,\pm \rangle
=h^{-1/2}\!\int_{-\infty }^{+\infty }\!dE \,e^{i(Et\mp \sqrt{2mE}\,X)/\hbar
}\,|E,\pm \rangle , 
\end{equation}
the states $|\tau ;X\rangle $ can be decomposed as a superposition of
negative- and positive-momentum contributions, in the form

\begin{equation}
\label{ec71}|\tau;X\rangle=|t\!=\!-\tau,-;X\rangle+|t\!=\!+\tau,+;X\rangle . 
\end{equation}

Inserting now the spectral resolution of $\hat {{\cal T}}(X)$, Eq. (\ref
{ec68}), into Eq. (\ref{ec62}) one can express the mean arrival time at the
spatial position $X$, in the form

\begin{equation}
\label{ec72}\langle t_X\rangle _{\pm }=\!\int_{-\infty }^{+\infty }\!d\tau
\,\left( \pm \tau \right) \,|\langle \tau ;X|\psi _{\pm }\rangle |^2 , 
\end{equation}
so that,

\begin{equation}
\label{ec64}\langle t_X\rangle_{+}=\!\int_{-\infty }^{+\infty }\!d\tau\,\tau
\,|\langle t\!=\!+\tau ,+;X|\psi _{+}\rangle |^2 , 
\end{equation}

\begin{equation}
\label{ec65}\langle t_X\rangle_{-}=\!\int_{-\infty }^{+\infty }\!d\tau \,
\left(-\tau \right) \,|\langle t\!=\!-\tau ,-;X|\psi _{-}\rangle |^2 . 
\end{equation}
On the other hand, taking into account the resolution of the unity,

\begin{equation}
\label{ec31b}\sum_{\alpha =\pm }\!\int_{-\infty }^{+\infty}\!dt\;|t,\alpha
;X\rangle \langle t,\alpha ;X|={\bf 1} , 
\end{equation}
we have

\begin{equation}
\label{ec66}1=\langle \psi _{\pm }|\psi _{\pm }\rangle =\!\int_{-\infty
}^{+\infty }\!dt\;|\langle t,\pm ;X|\psi _{\pm }\rangle |^2 , 
\end{equation}
which, for a free particle, coincides with the total probability of arriving
at the point $X$ at any instant.

Therefore, the quantities $|\langle \tau ;X|\psi _{\pm}\rangle |^2$
enter the above equations as a probability density, and lead to an
expression for the mean arrival time having the correct semiclassical limit
in terms of the probability current. However, unlike the latter, it is
definite positive. Accordingly, for a free particle in the Heisenberg state $%
|\psi _{+}\rangle $, one can interpret consistently $\langle\tau;X|\psi_{+}%
\rangle=\langle t\!=\!+\tau,+;X|\psi_{+}\rangle$ as the probability
amplitude of arriving at the spatial point $X$ from the left, at the instant 
$t=\tau $. Similarly, for a free particle in the Heisenberg state $|\psi
_{-}\rangle $, the scalar product $\langle -\tau ;X|\psi _{-}\rangle
=\langle t\!=\!+\tau ,-;X|\psi _{-}\rangle $ can be interpreted as the
probability amplitude of arriving at $X$ from the right, at the instant $%
t=\tau $.

\vspace{1.2 cm}
\begin{center}
{\large \bf 
{VI. PROBABILITY DISTRIBUTION OF ARRIVAL TIMES. \\ POTENTIAL BARRIER
\vspace{.2 cm}\\}}
\end{center}

Consider the passage of particles incident from the left over a
one-dimensional potential barrier $V(x)$. As usual, we assume that far away
from the scattering center, $V(x)$ vanishes sufficiently fast as to
guarantee the validity of the standard scattering theory formalism. Under
the conditions we are interested in, the ingoing asymptote, $|\psi _{{\rm in}%
}\rangle $, of the actual state of the particle satisfies

\begin{equation}
\label{psinpl}|\psi _{{\rm in}}\rangle \equiv \Theta (\hat{P})\,|\psi _{{\rm %
in}}\rangle . 
\end{equation}

The M\o ller operators $\hat \Omega _{\pm }$, which play a central role in
scattering theory, are defined as

\begin{equation}
\label{mol}\hat \Omega _{\pm }=\lim _{t\rightarrow \mp \infty }e^{i\hat
Ht/\hbar }\,e^{-i\hat H_0t/\hbar }, 
\end{equation}
where $\hat H_0\!=\!\hat P^2/2m$, and $\hat H=$ $\hat H_0+V(\hat X)$ is the
Hamiltonian governing the dynamical evolution of the system. These operators
have the importance that they map the asymptotic states onto the
corresponding scattering states. Specifically, the actual state of the
particle, $|\psi (t\!=\!0)\rangle $, is related to its in and out
asymptotes, $|\psi _{{\rm in}}\rangle $ and $|\psi _{{\rm out}}\rangle $, by
means of

\begin{equation}
\label{rela}|\psi (t\!=\!0)\rangle =\hat \Omega _{+}|\psi _{{\rm in}}\rangle
=\hat \Omega _{-}|\psi _{{\rm out}}\rangle . 
\end{equation}

Making use of the intertwining relations for the M\o ller operators [\ref
{Taylor}],

\begin{equation}
\label{ec67b}\hat \Omega _{\pm }^{\dagger }\,\hat H\,\hat \Omega _{\pm
}=\hat H_0, 
\end{equation}
the mean arrival time at a spatial point $X$, on the right of the barrier
and asymptotically far from the interaction region is given by

\begin{equation}
\label{asin}\langle t_X\rangle =\frac{\int_{-\infty }^{+\infty }d\tau \,\tau
\,\langle \psi (\tau )|\hat J(X)|\psi (\tau )\rangle }{\int_{-\infty
}^{+\infty }d\tau \,\langle \psi (\tau )|\hat J(X)|\psi (\tau )\rangle } , 
\end{equation}
where now we have


\begin{equation}
\label{ec68b}\langle \psi (\tau )|\hat{J}(X)|\psi (\tau )\rangle =\langle
\psi _{{\rm in}}|\,e^{i\hat{H}_0\tau/\hbar}\,\hat{\Omega}_{+}^{\dagger}\,%
\hat{J}(X)\,\hat{\Omega }_{+}e^{-i\hat{H}_0\tau /\hbar }\,|\psi_{{\rm in}%
}\rangle .
\end{equation}

Inserting twice the resolution of unity, Eq. (\ref{ec12}), and taking
advantage of Eq. (\ref{psinpl}) to write $\langle p|\psi _{{\rm in}}\rangle
=\Theta (p)\,\langle p|\psi _{{\rm in}}\rangle $, one obtains

\begin{equation}
\label{ec69b}\langle \psi (\tau )|\hat J(X)|\psi (\tau )\rangle
=\!\int_0^\infty \!\!dp^{\prime }\!\!\int_0^\infty\!\! dp\, e^{iE_{p^{\prime
}}\tau /\hbar}e^{-iE_p\tau /\hbar }\langle \psi_{{\rm in}}|p^{\prime
}\rangle \langle p^{\prime }|\hat \Omega_{+}^{\dagger }\,\hat J(X)\,\hat
\Omega _{+}|p\rangle \langle p|\psi _{{\rm in}}\rangle , 
\end{equation}
where $E_p=p^2/2m$.

The state $|p+\rangle \equiv \hat \Omega _{+}|p\rangle $, which is the
solution of the Lippmann-Schwinger equation corresponding to an ingoing
plane wave $|p\rangle $, satisfies the eigenvalue equation $\hat H\,|p+
\rangle=E_p\,|p+\rangle $ with the boundary conditions

\begin{equation}
\label{ec70b}x\rightarrow -\infty :\;\;\;\;\langle x|p+\rangle \sim \langle
x|p\rangle +R(p)\,\langle x|\!-\!p\rangle , 
\end{equation}

\begin{equation}
\label{ec71b}x\rightarrow +\infty :\;\;\;\;\langle x|p+\rangle \sim
T(p)\,\langle x|p\rangle , 
\end{equation}
$R(p)$ and $T(p)$ being the reflection and transmission coefficients,
respectively. Thus, for spatial points $X$ on the right and asymptotically
far from the interaction center one has

\begin{equation}
\label{ec72b}\langle p^{\prime}\!\!+\!|\hat J(X)|p+\rangle=T^{*}(p^{\prime
})\,T(p)\,\langle p^{\prime }|\hat J(X)|p\rangle ,
\end{equation}
where use has been made of Eqs. (\ref{curop}) and (\ref{ec71b}).
Substituting in (\ref{ec69b}) we obtain

\begin{equation}
\label{ec73}\langle \psi (\tau )|\hat J(X)|\psi (\tau )\rangle
=\!\!\int_0^\infty\!\! dp^{\prime }\!\!\int_0^\infty \!\!dp\,T^{*}
(p^{\prime })\langle\psi _{{\rm in}}|p^{\prime }\rangle \langle p^{\prime }
|e^{i\hat H_0\tau/\hbar }\hat J(X)\,e^{-i\hat H_0\tau /\hbar }|p\rangle
\langle p|\psi _{{\rm in}}\rangle T(p) . 
\end{equation}

Taking (\ref{ec73}) into account, the $\tau $-integral in the denominator of
(\ref{asin}) can be readily carried out to obtain a Dirac's delta,

\begin{equation}
\label{ec73bb}\delta (E_{p^{\prime }}-E_p)=\frac m{|p|}\,\delta (p^{\prime
}-p)+\frac m{|p|}\,\delta (p^{\prime }+p) ,
\end{equation}
and this can in turn be used to finally obtain

\begin{equation}
\label{transmi}\int_{-\infty }^{+\infty }\!d\tau \,\langle \psi (\tau )|\hat
J(X)|\psi (\tau )\rangle =\!\int_0^\infty \!dp\,|T(p)|^2\,|\langle p|\psi _{%
{\rm in}}\rangle |^2, 
\end{equation}
which is nothing but the transmittance. Defining the (unnormalized) freely
evolving transmitted state, $|\psi _{tr}\rangle $, as

\begin{equation}
\label{ec74}|\psi _{tr}\rangle \equiv\! \int_0^\infty \!dp\,T(p)\, \langle
p|\psi_{{\rm in}}\rangle \,|p\rangle , 
\end{equation}
and using Eq. (\ref{ec73}), the mean arrival time (\ref{asin}) takes the form

\begin{equation}
\label{ec75}\langle t_X\rangle =\frac 1{\langle \psi _{tr}|\psi _{tr}\rangle
}\,\langle \psi _{tr}|\!\int_{-\infty}^{+\infty}\!d\tau\,\tau\,\Theta (\hat
P)\,\hat J_{{\rm I}}(X,\tau )\,\Theta (\hat P)\,|\psi _{tr}\rangle , 
\end{equation}
where we have taken advantage of Eq. (\ref{psinpl}), and $\hat J_{{\rm I}%
}(X,\tau )$ denotes the current operator in the interaction picture,

\begin{equation}
\label{ec76}\hat J_{{\rm I}}(X,\tau )\equiv e^{i\hat H_0\tau /\hbar }\,\hat
J(X)\,e^{-i\hat H_0\tau /\hbar } . 
\end{equation}

A comparison between Eqs. (\ref{ec76}) and (\ref{ec40}) shows that the
current operator in the interaction picture coincides with the {\em free}
current operator in the Heisenberg picture. This is the important point
which allows us to rewrite $\langle t_X\rangle$ in terms of the freely
evolving operator $\hat {{\cal J}}_{+}(X)$ of Eq. (\ref{ec42}) and,
consequently exploit the formalism developed in the previous section for the
free case, to finally obtain

\begin{equation}
\label{ec77}\langle t_X\rangle =\frac{\langle \psi _{tr}|\hat {{\cal J}%
}_{+}(X)|\psi _{tr}\rangle }{\langle \psi _{tr}|\psi _{tr}\rangle }=\frac{%
\langle \psi _{tr}|\hat {{\cal T}}(X)|\psi _{tr}\rangle }{\langle \psi
_{tr}|\psi _{tr}\rangle } . 
\end{equation}
Inserting the expression for the self-adjoint operator $\hat {{\cal T}}(X)$,
given in Eqs. (\ref{ec68}) and (\ref{ec69}), the mean arrival time at a
spatial point $X$ behind the barrier and asymptotically far from the
interaction center, takes the suggestive form

\begin{equation}
\label{ec78}\langle t_X\rangle =\frac 1{\langle \psi _{tr}|\psi _{tr}\rangle
}\int_{-\infty }^{+\infty }\!d\tau \,\tau \,|\langle \tau ;X|\psi
_{tr}\rangle |^2. 
\end{equation}
Furthermore, taking into account that, for a given $X$, the states $\{|\tau
;X\rangle \}$ constitute a complete and orthonormal set, we find that the
integral

\begin{equation}
\label{ec79}\int_{-\infty }^{+\infty }\!d\tau \,|\langle \tau ;X|\psi
_{tr}\rangle |^2=\langle \psi _{tr}|\psi _{tr}\rangle =\!\int_0^\infty
\!dp\,|T(p)|^2\,|\langle p|\psi _{{\rm in}}\rangle |^2, 
\end{equation}
coincides with the transmittance, which is nothing but the total probability
of arriving at an asymptotic point behind the barrier. Therefore, we can
consistently interpret $\langle \tau ;X|\psi _{tr}\rangle =\langle
t\!=\!\tau ,+;X|\psi _{tr}\rangle $ as the (unnormalized) probability 
amplitude of arriving at the asymptotic point $X$, behind the barrier, at 
the instant $t=\tau $.

The above results can be expressed in terms of the ingoing asymptote $|\psi
_{{\rm in}}\rangle $, by using the scattering operator ${\hat S}\equiv \hat
\Omega_{-}^{\dagger }\,\hat \Omega _{+}$, which relates the in and out
asymptotes, $|\psi _{{\rm out}}\rangle ={\hat S}\,|\psi_{{\rm in}}\rangle$.
Indeed, it is shown in the Appendix that the freely evolving transmitted state 
$|\psi _{tr}\rangle $ can be written as

\begin{equation}
\label{ec79b}|\psi _{tr}\rangle =\Theta (\hat P)\,{\hat S}\,|\psi _{{\rm in}%
}\rangle , 
\end{equation}
so that the (unnormalized) probability density of arriving at an asymptotic 
point $X$, behind the barrier, at the instant $t=\tau $ reads

\begin{equation}
\label{ec79c}|\langle \tau ;X|\,\Theta (\hat P)\,{\hat S}\,|\psi_{{\rm in}}
\rangle|^2=|\langle t\!=\!\tau ,+;X|\,{\hat S}\,|\psi_{{\rm in}}\rangle |^2 . 
\end{equation}

Finally, it should be noted that when the wave packet corresponding to the
actual scattering state at $t=0$ (which is assumed to be a linear
superposition of only positive plane waves) does not overlap with the
potential barrier, then it becomes physically indistinguishable from the
asymptotic ingoing wave packet and the above equations hold true with the
substitution $|\psi _{{\rm in}}\rangle \rightarrow |\psi (t\!=\!0)\rangle $.

\vspace{1.2 cm}
\begin{center}
{\large \bf {VII. TIME--ENERGY UNCERTAINTY RELATION\vspace{.2 cm}\\}}
\end{center}

Giving a precise meaning to the well-known time--energy uncertainty relation
seems to be a reasonable requirement for any quantum formulation of the
time-of-arrival concept.

As already said, the commutation relation $[\hat H,\hat T]=i\hbar $
automatically leads to the uncertainty relation (\ref{ec1pb}). Although, it
has not been possible to develop a quantum formulation of the arrival-time
concept based on such a commutation relation, there is still room for a
time--energy uncertainty relation, because even though the existence of a
commutation relation is a sufficient condition for the existence of an
uncertainty relation, it is by no means a necessary condition.

It should be noted that the self-adjoint operator $\hat {{\cal T}}(X)$
defined by Eqs. (\ref{ec67})--(\ref{ec69}) is conjugate to the operator $%
\hat {{\cal H}}$. Specifically,

\begin{equation}
\label{ec80}[\hat {{\cal H}},\hat {{\cal T}}(X)]=e^{-i\hat PX/\hbar }\,[\hat
{{\cal H}},\hat {{\cal T}}]\,e^{+i\hat PX/\hbar }=i\hbar . 
\end{equation}

Introducing a probability amplitude for the time of arrival, in terms of the
eigenvalues and eigenstates of a self-adjoint operator satisfying the above
commutation relation has as an important consequence the existence of a
time--energy uncertainty relation. To see this, let us consider the problem
studied in the previous section, namely, the arrival time of a quantum
particle at a detector located behind a potential barrier and asymptotically
far from the interaction center. [The free case is nothing but a particular
case of the latter corresponding to $\,T(p)\rightarrow 1$, which implies $%
|\psi_{tr}\rangle \rightarrow |\psi _{{\rm in}}\rangle $.]

Because of Eq. (\ref{ec80}), it automatically holds

\begin{equation}
\label{ec81}\Delta {\cal H}\;\Delta {\cal T}_X\geq \hbar /2 , 
\end{equation}
where $\Delta {\cal H}\,$ and $\Delta {\cal T}_X$ are the root-mean-square
deviations of the corresponding observables, i.e., $(\Delta {\cal H}%
)^2\equiv \langle \hat {{\cal H}}^2\rangle -\langle \hat {{\cal H}}\rangle
^2 $, and $(\Delta {\cal T}_X)^2\equiv \langle \hat {{\cal T}}^2(X)\rangle
-\langle \hat {{\cal T}}(X)\rangle ^2$, with $\langle \hat {{\cal A}}\rangle
\equiv \langle \psi _{tr}|\hat {{\cal A}}|\psi _{tr}\rangle /\langle \psi
_{tr}|\psi _{tr}\rangle $. However, from Eqs. (\ref{ec18}) and (\ref{ec74})
it follows that

\begin{equation}
\label{ec82}\hat {{\cal H}}\,|\psi _{tr}\rangle =\hat H_0\,|\psi
_{tr}\rangle , 
\end{equation}
and hence $\Delta {\cal H}$ coincides with the statistical spread of the
energy of particles arriving at the detector,

\begin{equation}
\label{ec83}(\Delta {\cal H})^2=\langle \hat H_0^2\rangle -\langle \hat
H_0\rangle ^2\equiv (\Delta E)^2 . 
\end{equation}

On the other hand, according to Eqs. (\ref{ec68}) and (\ref{ec77}) we have 
\begin{equation}
\label{ec84}(\Delta {\cal T}_X)^2=\langle \left( \hat {{\cal T}}(X)-\langle
t_X\rangle \right) ^2\rangle =\frac 1{\langle \psi _{tr}|\psi _{tr}\rangle
}\int_{-\infty }^{+\infty }\!d\tau \,(\tau -\langle t_X\rangle )^2\,|\langle
\tau ;X|\psi _{tr}\rangle |^2\equiv (\Delta t_X)^2 , 
\end{equation}
and since $\,|\langle \tau ;X|\psi _{tr}\rangle |^2/\langle \psi _{tr}|\psi
_{tr}\rangle $ is the probability distribution of the arrival time of
particles at the detector, the above equation shows that $\Delta {\cal T}_X$
is nothing but the corresponding statistical deviation, $\Delta t_X$.
Therefore, the statistical spreads of the energy, $\Delta E$, and time of
arrival, $\Delta t_X$, of particles reaching the detector satisfy the
time--energy uncertainty relation

\begin{equation}
\label{ec85}\Delta E\;\Delta t_X\geq \hbar /2 .
\end{equation}

\vspace{1.2 cm}
\begin{center}
{\large \bf {VI. CONCLUSION\vspace{.2 cm}\\}}
\end{center}

Despite its fundamental nature, the quantum formulation of the
time-of-arrival concept is a problem which remains open nowadays. This
question has the additional interest that probability distributions of
arrival times are, in principle, experimentally accessible via the
time-of-flight technique. Moreover, a quantum formulation of such a problem
may provide a useful tool for a better understanding of the tunneling time
problem as well as its possible technological applications.

The main difficulty for defining a quantum time operator lies in the
nonexistence, in general, of a self-adjoint operator conjugate to the
Hamiltonian, a problem which can be always traced back to the semibounded
nature of the energy spectrum. In turn, the lack of a self-adjoint time
operator implies the lack of a properly and unambiguously defined
probability distribution of arrival times.

Although it has been shown that under certain circumstances of physical
interest, the probability current becomes positive and admits a proper
interpretation as an unnormalized probability distribution of the time of
arrival [\ref{Leav2}--\ref{Mug}], it cannot be considered as a fully
satisfactory solution of the problem, for it is not positive definite as it
should be.

In searching for a probability distribution defined through a quantum time
operator one has to circumvent the problem stated above. There are two
possibilities. If one decides that any proper time operator must be strictly
conjugate to the Hamiltonian, then one has to renounce to find a
self-adjoint operator. (Even though such a property is a hallmark of any
observable in the standard quantum formalism, it is not strictly necessary
for a consistent formulation of probability distributions of measurements
results [\ref{Busch},\ref{Gian}].) If, on the contrary, one imposes
self-adjointness as a desirable requirement for any observable, then one
necessarily has to abandon the requirement that such an operator be
conjugate to the Hamiltonian. In the present paper we have adopted the
latter via. We have explicitly constructed a self-adjoint operator $\hat {%
{\cal H}}$ with dimensions of energy, and a non-bounded spectrum. Such an
operator is essentially the energy of the particle with the sign of its
momentum. The non-bounded character of its spectrum enables us to introduce
a self-adjoint operator with dimensions of time, $\hat {{\cal T}}$, by
demanding it to be conjugate to $\hat {{\cal H}}$. Since the latter is
essentially the Hamiltonian, except sign, one expects the self-adjoint
operator $\hat {{\cal T}}$ so defined to be physically meaningful and
relevant to the arrival time problem. Indeed, we have shown that it is
possible to define consistently a probability distribution of arrival times
at a spatial point, in terms of the eigenvalues and eigenstates of such an
operator. This probability distribution, which is a function of the initial
state of the system, does not depend on the particular design of the
measuring device, and has the additional desirable consequence that it leads
to a precisely defined time--energy uncertainty relation.

\vspace{1.2 cm}
\begin{center}{\large \bf 
{APPENDIX: TRANSMITTED STATE AS A FUNCTION OF THE INGOING ASYMPTOTE
\vspace{.2 cm}\\}}
\end{center}

In this appendix we show that the freely evolving transmitted state $|\psi
_{tr}\rangle $ can be written in terms of the ingoing asymptote $|\psi _{%
{\rm in}}\rangle $ in the form [Eq. (\ref{ec79b})]

\begin{equation}
\label{eca1}|\psi _{tr}\rangle =\Theta (\hat P)\,|\psi _{{\rm out}}\rangle
=\Theta (\hat P)\,{\hat S}\,|\psi _{{\rm in}}\rangle ,
\end{equation}
where ${\hat S}$ denotes the scattering operator, relating the ingoing and
outgoing asymptotes,

\begin{equation}
\label{eca2}
|\psi _{{\rm out}}\rangle ={\hat S}\,|\psi_{{\rm in}}\rangle =\!\int_{-\infty
}^{+\infty }\!\!dp^{\prime }\!\!\int_{-\infty }^{+\infty }\!\!dp\,|p^{\prime
}\rangle \langle p^{\prime }|\,{\hat S}\,|p\rangle \langle p|\psi _{{\rm in}%
}\rangle . 
\end{equation}

Taking into account that the matrix elements of the scattering operator can be 
written in terms of the {\em on-the-energy-shell\,} $\hat{\sf T}$ {\em matrix}
as [\ref{Taylor}]

\begin{equation}
\label{eca3}
\langle p^{\prime }|\,{\hat S}\,|p\rangle =\delta (p-p^{\prime })-2\pi
i\,\delta (E_p-E_{p^{\prime }})\,\langle p^{\prime }|\,\hat{\sf T}%
(E_p+i0)\,|p\rangle , 
\end{equation}
equation (\ref{eca2}) reads

\begin{eqnarray}
\label{eca4}
|\psi_{{\rm out}}\rangle ={\hat S}\,|\psi_{{\rm in}}\rangle \!
&=& \!\int_0^\infty\!\!dp\,\left( 1-2\pi im/p\,\langle p|\,
    \hat{\sf T}(E_p+i0)\,|p\rangle \right) \langle p
    |\psi _{{\rm in}}\rangle \, |p\rangle + \nonumber \\
& & \!\int_0^\infty \!\!dp\,\left(
-2\pi im/p\,\langle -p|\,\hat{\sf T}(E_p+i0)\,|p\rangle \right) 
\langle p|\psi _{{\rm in}}\rangle \mid\!\! -p \rangle . 
\end{eqnarray}

On the other hand, from the Lippmann-Schwinger equation for $|p+\rangle $ it
follows that the wave function $\langle x|p+\rangle $ can be written as

\begin{equation}
\label{eca5}
\langle x|p+\rangle =\langle x|p\rangle +\!\int\! dx^{\prime }\langle 
x|\, (E_p+i0-\hat H_0)^{-1}|x^{\prime }\rangle \langle x^{\prime }|\,
\hat{\sf T} (E_p+i0)\,|p\rangle .
\end{equation}
Substituting in the above equation the expression for the Green's 
function (which can be easily obtained by inserting the resolution of 
unity in terms of momentum eigenstates and evaluating the resulting 
integral by contour integration in the complex plane),

\begin{equation}
\label{eca6}
\langle x|\, (E_p+i0-\hat H_0)^{-1}|x^{\prime }\rangle =-\frac{im}{%
\hbar |p|}e^{i|p|\,|x-x^{\prime }|/\hbar } ,
\end{equation}
one obtains, for $p>0$ and $x\rightarrow +\infty $,

\begin{eqnarray}
\label{eca7}
\langle x|p+\rangle &\sim& \langle x|p\rangle -2\pi im/p\,\langle x|p\rangle
\int\! dx^{\prime }\langle p|x^{\prime }\rangle \langle x^{\prime }|\,
\hat{\sf T} (E_p+i0)\,|p\rangle = \nonumber \\
& &\left( 1-2\pi im/p\,\langle p|\,\hat{\sf T} (E_p+i0)\,|p\rangle \right) 
\,\langle x|p\rangle .
\end{eqnarray}
A comparison with Eq. (\ref{ec71b}) yields

\begin{equation}
\label{eca9}
T(p)=\left( 1-2\pi im/p\,\langle p|\,\hat{\sf T}(E_p+i0)\,|p\rangle \right) .
\end{equation}
Similarly, for $p>0$ and $x\rightarrow -\infty$, Eq. (\ref{eca5}) leads to

\begin{eqnarray}
\label{eca10}
\langle x|p+\rangle &\sim& \langle x|p\rangle -2\pi im/p\,\langle x
\mid\!\! -p \rangle
\int\! dx^{\prime }\langle -p|x^{\prime }\rangle \langle x^{\prime }|\,
\hat{\sf T} (E_p+i0)\,|p\rangle = \nonumber \\
& &\langle x|p\rangle +\left( -2\pi im/p\,\langle -p|\, \hat{\sf T}
(E_p+i0)\,|p\rangle \right) \langle x|p-\rangle ,
\end{eqnarray}
and comparing with Eq. (\ref{ec70b}) we find

\begin{equation}
\label{eca12}
R(p)=\left( -2\pi im/p\,\langle -p|\,\hat{\sf T}(E_p+i0)\,|p\rangle \right) . 
\end{equation}
Substituting Eqs. (\ref{eca9}) and (\ref{eca12}) in (\ref{eca4}) we finally 
arrive at

\begin{equation}
\label{eca13}
|\psi _{{\rm out}}\rangle ={\hat S}\,|\psi _{{\rm in}}\rangle
=\!\int_0^\infty \!dp\,T(p)\,\langle p|\psi _{{\rm in}}\rangle \,|p\rangle
+\!\int_0^\infty \!dp\,R(p)\,\langle p|\psi _{{\rm in}}\rangle 
\mid\!\! -p \rangle ,
\end{equation}
from which it follows Eq. (\ref{eca1}).

\newpage
\vspace{1.4 cm}
\begin{center}
{\large \bf  REFERENCES \vspace{.6cm}}
\end{center}

\begin{enumerate}
\item  \label{Raz} M. Razavy, Am. J. Phys. {\bf 35}, 955 (1967); Nuovo
Cimento B {\bf 63}, 271 (1969).

\item  \label{Ko} D. H. Kobe, Am. J. Phys. {\bf 61}, 1031 (1993).

\item  \label{QF} see for example, N. N. Bogoliubov and D. V. Shirkov,
{\it Quantum Fields} (Benjamin/Cummings, Reading, Massachusetts, 1983).

\item  \label{Pauli} W. Pauli, in {\it Encyclopaedia of Physics},
 edited by S. Flugge, Vol. 5/1, p. 60 (Springer, Berlin, 1958).

\item  \label{Aharo} Y. Aharonov and D. Bohm, Phys. Rev. {\bf 122}, 1649
(1961).

\item  \label{Lan} M. B\"uttiker and R. Landauer, Phys. Rev. Lett. {\bf 49},
1739 (1982).

\item  \label{Rev} For recent reviews on the subject see (a) E. H. Hauge and
J. A. St\/ovneng, Rev. Mod. Phys. {\bf 61}, 917 (1989); (b) M. B\"uttiker,
in {\it Electronic Properties of Multilayers and Low-Dimensional
Semiconductor Structures}, edited by J. M. Chamberlain {\it et al.}, p. 297,
(Plenum, New York, 1990); (c) R. Landauer, Ber. Bunsenges. Phys. Chem. {\bf %
95}, 404 (1991); (d) C. R. Leavens and G. C. Aers, in {\it Scanning
Tunneling Microspy III}, edited by R. Wiesendanger and H. J. G\"utherodt,
pp. 105--140 (Springer, Berlin, 1993); (d) R. Landauer and T. Martin, Rev.
Mod. Phys. {\bf 66}, 217 (1994).

\item  \label{Bu} M. B\"uttiker, Phys. Rev. B {\bf 27}, 6178 (1983).

\item  \label{Du} R. S. Dumont and T. L. Marchioro II, Phys. Rev. A {\bf 47}%
, 85 (1993).

\item  \label{San} S. Brouard, R. Sala, and J. G. Muga, Phys. Rev. A {\bf 49}%
, 4312 (1994).

\item  \label{Reca} V. S. Olkhovsky and E. Recami, Phys. Rep. {\bf 214}, 339
(1992).

\item  \label{Leav} C. R. Leavens, Solid State Commun. {\bf 85}, 115 (1993); 
{\bf 89}, 37 (1993).

\item  \label{VDB1} V. Delgado, S. Brouard, and J. G. Muga, Solid State
Commun. {\bf 94}, 979 (1995).

\item  \label{Low} F. E. Low and P. F. Mende, Ann. Phys. (N.Y.) {\bf 210},
380 (1991).

\item  \label{VDB2} V. Delgado and J. G. Muga, Ann. Phys. (N.Y.) {\bf 248},
122 (1996).

\item  \label{Allco} G. R. Allcock, Ann. Phys. (N.Y.) {\bf 53}, 253 (1969); 
{\bf 53}, 286 (1969); {\bf 53}, 311 (1969).

\item  \label{Kij} J. Kijowski, Rep. Math. Phys. {\bf 6}, 361 (1974).

\item  \label{Leav2} C. R. Leavens, Phys. Lett. A {\bf 178}, 27 (1993).

\item  \label{Leav3} W. R. McKinnon and C. R. Leavens, Phys. Rev. A {\bf 51}%
, 2748 (1995).

\item  \label{Mug} J. G. Muga, S. Brouard, and D. Mac\'\i as, Ann. Phys.
(N.Y.) {\bf 240}, 351 (1995).

\item  \label{Grot} N. Grot, C. Rovelli, and R. S. Tate, Phys. Rev. A
{\bf 54}, 4676 (1996).

\item  \label{Jleon} J. Le\'on, ''Time-of-arrival formalism for the
relativistic particle'', preprint quant-ph/9608013 (1996).

\item  \label{Mielnik} B. Mielnik, Found. Phys. {\bf 24}, 1113 (1994).

\item  \label{Piron} C. Piron, C. R. Acad. Seances (Paris) A {\bf 286}, 
713 (1978).

\item  \label{Wig2} F. Goldrich and E. P. Wigner, in {\it Magic without 
Magic: John Archibald Wheeler}, edited by J. R. Klauder, p. 147 (W. H. 
Freeman, San Francisco, 1972).

\item  \label{Taylor} J. R. Taylor, {\it Scattering Theory: The Quantum 
Theory on Nonrelativistic Collisions} (John Wiley \& Sons, Inc., New York, 
1975).

\item  \label{Busch} P. Busch, M. Grabowski, and P. J. Lahti, Phys. Lett. A 
{\bf 191}, 357 (1994).

\item  \label{Gian} R. Giannitrapani, ''On a Time Observable in Quantum
Mechanics'', preprint quant-ph/9611015 (1996).
\end{enumerate}

\end{document}